\documentstyle[aps, preprint]{revtex}
\draft
\begin{document}
\title{Effective Hamiltonian Approach to the Master Equation}
\author{X.X. Yi and S.X. Yu}
\address{The Institute of Theoretical Physics, Academia Sinica, P.O.Box
2735, Beijing 100080, China}
\date{\today}
\maketitle
\begin{abstract}
A method of exactly solving the master equation is presented in
this letter. The explicit form of the solution is determined by
the time evolution of a composite system including an auxiliary
system and the open system in question. The effective Hamiltonian
governing the time evolution of the composed system are derived
from the master equation. Two examples, the dissipative two-level
system and the damped harmonic oscillator, are presented to
illustrate the solving procedure.
\end{abstract}

The problem of open quantum systems has been around since the
beginnings of quantum mechanics [1]. Important contributions to
this general area have been made by researchers working in fields
as diverse as cosmology [2], condensed matter [3], quantum optics
[4,5], particle physics [6], quantum measurement [7], and quantum
computation [8]. The problem can be described generally as
interest in the effective dynamics of the open system surrounded
by its environment. A formal framework to describe the effective
dynamics of such an open system was set up in Ref.[9].

Generally speaking, interactions between a quantum system and its
environment result in two kinds of irreversible effects:
dissipation and decoherence. The first effect is due to the energy
exchange between the system and its environment, whereas the
second one comes from the system-environment interaction that does
not change the system energy. A powerful tool to study the quantum
dissipative system is the master equation, which can be obtained
in Markovian limit [1,2,4]. This approximation is often very
useful became it is valid for many physically relevant situations
and became its numerical solutions can be found. As given by
Gardiner, Walls and Millburn, Louisell in their textbook [4], the
reduced density matrix $\rho$ of the open system which is linearly
coupled to its environment obeys the following master equation in
Lindblad form [10]
\begin{eqnarray}
\dot{\rho}(t)&=&-i[H_0,\rho]\cr &+&\frac 1 2 \sum_m K_m(2X_m^-\rho
X^+_m- X_m^+X_m^-\rho-\rho X_m^+X_m^-)\nonumber\\ &+&\frac 1 2
\sum_mG_m(2X_m^+\rho X^-_m- X_m^-X_m^+\rho-\rho X_m^-X_m^+)
\end{eqnarray}
with $$K_m=2 \mbox{Re}\left[\int_0^{\infty}d \tau e^{i\omega_m
\tau}{\rm Tr}_{env}\{A_m(\tau) A^{\dag}_m(0)\rho_{env}\}\right],$$
$$G_m=2 \mbox{Re}\left[\int_0^{\infty}d \tau e^{i\omega_m
\tau}{\rm Tr}_{env}\{A_m^{\dag}(\tau) A_m(0)\rho_{env}\}\right].$$
Here, $\rho(t)=\rho(t,K_m,G_m)$ stands for the density operator of
the system and $\rho_{env}$ denotes the density operator of the
environment, $X_m^{\pm}$ are eigenoperators of the system
satisfying $[H_0,X^{\pm}_m]=\pm\hbar\omega_m X_m^{\pm}$,
 $H_0$ stands for the free Hamiltonian of the system, and
 $A_m$($A_m^{\dagger}$) are operators of the environment through
 which the system and its environment coupled together.
 Notice from eq.(1) that $G_m$ should vanish at zero temperature
$T=0$, while $K_m$ should not if $A_m$ are indeed destruction
operators of some kind. Some efforts such as the short-time
expansion[11], the exact solution for some special kinds of the
master equation[12], the small lose rate expansion[13] and the
method of stochastic unravelings[14] have been made to solve (or
solve it alternatively) the master equation. Unfortunately, this
kind of master equation has not a general exact analytical
solution yet.

Here we shall construct an exact solution for the master equation
to describe quantum dynamics of an open system with linear
dissipation. The method is outlined as follows. After introducing
an auxillary system and mapping the density matrix of the original
system to a pure state of the composite system, we obtain an
effective Hamiltonian describing the evolution of the composite
system from the master equation satisfied by the density matrix.
Then the solution of the master equation can be obtained in terms
of the evolution of the composite system by mapping the pure state
back to the density matrix. As its applications, we present two
examples to illustrate the solving procedure. The first example is
a two-level atom coupling to a bosonic environment, whereas the
second example consists of a single mode field in a cavity with
linewidth $\kappa$ due to partial transmission through one mirror.

We suppose that the Hilbert space of the system $S$ is an
$N$-dimension Hilbert space spanned by, for instance, all the
eigenstates of $H_0$. To begin with, we rewrite the master
equation given in eq.(1) as the following general form
 \begin{equation}
 i\frac{\partial \rho}{\partial t} =H\rho -\rho H^{\dagger}
 +i\sum_\alpha \gamma_\alpha L_\alpha\rho L_\alpha^{\dagger},
 \end{equation}
where $H$ and $L_\alpha$ are well defined time-independent
operators of the system, which may not be Hermitian generally and
$\gamma_\alpha$ are real parameters determined by the environment.

As the first step to solve this general master equation, we
introduce an auxiliary system $A$ which is the same with the
original system we concern. That is to say, the original Hilbert
space is extended to an $N^2$-dimensional Hilbert space, which is
the Hilbert space of the original system $S$ and the ancilla $A$.
Let $\{|m_S\rangle |n_A\rangle\}$ be an orthonormal and complete
basis for this composite system.

A density matrix $\rho$ of the system $S$, whose matrix elements
under the basis $\{|m_S\rangle\}$ of the system are denoted as
$\rho_{mn}=\langle m|\rho|n\rangle_S$,  determines a pure
bipartite state in the $N^2$-dimensional Hilbert space according
to
\begin{equation}
|\psi_{\rho}\rangle=\sum_{m,n=1}^{N} \rho_{mn}
|m_S\rangle|n_A\rangle.
\end{equation}
We note that this corresponding pure bipartite state is generally
not normalized unless the initial state $\rho$ of the system is a pure
state. In fact every operator of the original system determines a
pure state of the bipartite system in the same manner \cite{pm}.

With the density matrix $\rho$ evolving with time, the
corresponding pure state $|\psi_{\rho}\rangle$ changes accordingly
into another pure state
\begin{equation}\label{def}
|\psi_{\rho}(t)\rangle=\sum_{m,n=1}^{N}\rho_{mn}(t)
|m_S\rangle|n_A\rangle.
\end{equation}
Because the evolution of the density matrix is governed by the
general master equation (2), the evolution of the pure state
$|\psi_{\rho}(t)\rangle$ is governed by the following
Sch\"odinger-like equation
\begin{equation}
i\partial_t |\psi_{\rho}(t)\rangle=\tilde{H}
|\psi_{\rho}(t)\rangle.
\end{equation}
Here the effective Hamiltonian, which is generally not Hermitian,
reads
\begin{equation}
\tilde{H}=H -{H}_A +i\sum_\alpha\gamma_\alpha L_{\alpha}
{L}_{A\alpha},
\end{equation}
where operators $H$ and $L_\alpha$ of the original system are the same
as given in Eq.(2) and operators $H_A$ and ${L}_{A\alpha}^*$ are
operators of the auxiliary system whose matrix elements are
specified by
\begin{eqnarray}
&\langle m|{H}_A|n\rangle_A=\langle
 n|H^{\dagger}|m\rangle_S, &\\
& \langle m |{L}_{A\alpha}|n\rangle_A=\langle
n|L_\alpha^{\dagger}|m\rangle_S.&
\end{eqnarray}
We note that the first two terms of the effective Hamiltonian
$\tilde H$ describe free evolutions of the system and the
auxiliary system governed by $H$ and $-H_A$ and the third term
describes an interaction between the system and  the ancilla.

Because the effective Hamiltonian $\tilde H$ of the composite
system  is time-independent, we can obtain formally the evolution
of the pure state $|\psi_{\rho}(t)\rangle$ as
 \begin{equation}
 |\psi_{\rho}(t)\rangle=e^{-i\tilde H t}|\psi_{\rho}\rangle.
 \end{equation}
>From the definition Eq.(4) of the pure state
$|\psi_{\rho}(t)\rangle$, we obtain finally the time-dependence of
the original density matrix as
\begin{equation}\label{sol}
 \rho_{mn}(t)=\langle m_S|\langle
 n_A|\psi_{\rho}(t)\rangle=\langle m_S|\langle n_A|e^{-i\tilde H
 t}|\psi_{\rho}\rangle.
\end{equation}
Since the effective Hamiltonian $\tilde H$  changes its sign under
the complex conjugate together with an interchange of the system
and the ancilla, the Hermicity of the density matrix is preserved
during the evolution. Moreover if the master equation (2)
preserves the trace of the density matrix, solution given in
Eq.(10) has also trace 1 when the initial state is normalized.

The generalizations of the above effective Hamiltonian method to
the systems with infinite many energy levels or with continuous
spectra are straightforward. Also, the master equations with
time-dependent $\gamma_\alpha$ and those master equations not
posessing Lindblad form can be treated in the same manner, i.e.,
effective Hamiltonian can be obtained similarly. Thus the problem
of solving the master equation becomes a problem of finding the
time evolution of an effective Hamiltonian. Although the
biorthogonal basis can be used to deal with general non-Hermitian
Hamiltonian, for some special cases we can evaluate the evolution
operator directly for finite-level system as illustrated by the
first example and factorize the evolution operator directly  for
quadratic systems as illustrated by the second example.

In order to gain further insight into the content of the solution
presented above, we shall consider at first the following master
equation
describing a two-level atom coupled to a
bose-mode environment [4]
\begin{equation}
\dot{\rho}=-\frac{i\Omega}2 [\sigma_z,\rho]+\frac \gamma 2 \{
2\sigma_-\rho\sigma_+-\rho\sigma_+\sigma_--\sigma_+\sigma_-\rho\}
\end{equation}
with $$\gamma=2\pi \mbox{Re}\left[\int_0^{\infty}d\tau
e^{i\omega_m\tau} {\rm Tr}_{env}\{b_m(\tau)
b^{\dag}_m(0)\rho_{env}\}\right],$$ where $b_m^{\dag} (b_m)$
stands for the creation (annihilation) operator of the $m$-th mode
of the environment, $\Omega$ is the Rabi frequency, and $\sigma_z
(\sigma_\pm)$ denote the Pauli matrices. The basis of the
two-level system is chosen so that
$\sigma_+|g_S\rangle=|e_S\rangle$ where $|g_S\rangle$ and
$|e_S\rangle$ stand for the ground and excited states. This master
equation is obtained under the condition that the environment is
in its vacuum state.

The auxiliary system we shall introduce is another two-level
system whose Pauli matrices are denoted as $\tau_z (\tau_\pm)$
which satisfy $\tau_+|g_A\rangle=|e_A\rangle$, where $|e_A\rangle$
denotes the upper level of the ancilla  A. By rewriting the master
equation above into the general form of Eq.(2), we obtain
immediately $H=(\Omega\sigma_z-i\gamma\sigma_+\sigma_-)/2$ and
$L=\sigma_-$. According to the definitions (7-8) we obtain
$H_A=(\Omega\tau_z+i\gamma\tau_+\tau_-)/2$ and $L_A=\tau_-$. The
effective Hamiltonian governing the time evolution of the
bipartite system is therefore
\begin{eqnarray}
\tilde{H}&=&H-H_A
+i\gamma\sigma_-\tau_-.
\end{eqnarray}
As mentioned above, eq.(12) is the effective Hamiltonian
corresponding to the master eq.(11), which is derived by
neglecting the temperature effect of the environment. If we take
the temperature effects into account, the effective Hamiltonian
should be
\begin{eqnarray}
\tilde{H}_T&=&H-H_A+i\gamma\left(\bar{N}\sigma_+\tau_+
+(\bar{N}+1)\sigma_-\tau_--\bar{N}\right)\cr
&=&\left\{\frac\Omega2(\sigma_z-\tau_z)-i\gamma\left(\bar
N+\frac12\right)\right\}\cr&&+i\gamma \left\{\bar{N}\sigma_+\tau_+
+(\bar{N}+1)\sigma_-\tau_--\frac{\sigma_z+\tau_z}4\right\}\cr
&\equiv& H_0+i\gamma J
\end{eqnarray}
where $\bar{N}=(\exp(\Omega/k_BT)-1)^{-1}$ is the Bose
distribution. Further we have $[H_0,J]=0$ and
$J|e_S\rangle|g_A\rangle=J|g_S\rangle|e_A\rangle=0$ which are
sufficient to make an explicit calculation of the evolution
operator
 \begin{equation}
 e^{-i\tilde H_Tt}=\frac1{2\bar N+1}\left(
 \matrix{\bar N+(\bar N+1)e^{-(2\bar N+1)\gamma t}
 &0&0&\bar N(1-e^{-(2\bar N+1)\gamma t})\cr
 0&e^{-i\Omega t-(\bar N+1/2)\gamma t}&0&0\cr
 0&0&e^{i\Omega t-(\bar N+1/2)\gamma t}&0\cr
 (\bar N+1)(1-e^{-(2\bar N+1)\gamma t})&0&0&\bar N+1+\bar Ne^{-(2\bar
N+1)\gamma t}\cr}
 \right)
 \end{equation}
where the bases of the composite system has been arranged as
$|e_S,e_A\rangle$, $|e_S,g_A\rangle$, $|g_S,e_A\rangle$, and
$|g_S,g_A\rangle$. For simplicity, we consider zero temperature
Hamiltonian $\tilde{H}$ to study the time evolution of the density
matrix $\rho(t)$ with $\bar N=0$.

For a general initial state
$\rho(0)=\sum_{i,j=g,e}\rho_{ij}|i_S\rangle\langle j_S|$ where
$\rho_{ij}=\langle i_S|\rho(0)|j_S\rangle$, the corresponding
initial state of the bipartite system  is
$|\psi_{\rho}(0)\rangle=\sum_{i,j=g,e}\rho_{ij}|i_S,j_A\rangle$.
With this initial condition the  final state of the composed
system at time $t$ reads
\begin{eqnarray}
|\psi_{\rho}(t)\rangle&=&\rho_{ee} e^{-\gamma t}
|e_S,e_A\rangle+(\rho_{gg}+\rho_{ee} (1-e^{-\gamma t})
)|g_S,g_A\rangle \nonumber\\ &&+\rho_{eg}
e^{-\frac{\gamma}{2}-i\Omega t}|e_S,g_A\rangle+\rho_{ge}
e^{-\frac{\gamma}{2}+i\Omega t}|g_S,e_A\rangle.
\end{eqnarray}
That is
\begin{eqnarray}
\rho_{ee}(t)&=&\rho_{ee}e^{-\gamma t}, \ \ \rho_{gg}(t)=\rho_{ee}
(1-e^{-\gamma t})+\rho_{gg},\nonumber\\
\rho_{eg}(t)&=&e^{-\frac{\gamma}{2}t-i\Omega t}\rho_{eg},\ \
\rho_{ge}(t)=e^{-\frac{\gamma}{2}t+i\Omega t}\rho_{ge}.
\end{eqnarray}
The results show that the off-diagonal elements of the density
matrix are damped-oscillation function of time, while the diagonal
element $\rho_{ee}$ decay exponentially. If $\Omega>\gamma$, there
are several Rabi oscillations in $\rho_{eg}$ (or $\rho_{ge}$) with
the time evolution, otherwise $\rho_{eg}$ (or $\rho_{ge}$) decay
directly. Especially we consider a state $\rho=|e_S\rangle\langle
e_S|$ as the initial condition of the density operator $\rho$, At
time $t$ we have
\begin{eqnarray}
\langle e_S|\rho(t)|e_S\rangle=e^{-\gamma t},\langle
g_S|\rho(t)|g_S\rangle=1-e^{-\gamma t},\nonumber\\ \langle
e_S|\rho(t)|g_S\rangle=\langle g_S|\rho(t)|e_S\rangle =0.
\end{eqnarray}
It is well known that the element $\langle e_S|\rho(t)|e_S\rangle$
of the density operator $\rho(t)$ represents the population of the
system in its upper level $|e_S\rangle$. The results (17) show
that the population of the upper level $|e_S\rangle$ decay
exponentially with the time evolution, this coincides with the
results given in most textbooks [4].

The second example presented here is a single-mode field in a
lossy cavity. The density operator for that mode obeys the
following master equation in the Schr\"odinger picture[4],
\begin{equation}
\dot{\rho}=-i[\omega_f a^{\dag}a,\rho]+\frac{\kappa}{2}(2a\rho
a^{\dag}- a^{\dag}a\rho- \rho a^{\dag}a),
\end{equation}
where $\kappa$ is the linewidth of the cavity mode with frequency
$\omega_f$. In most textbooks, the solution of the master equation
is given in terms of diagonal matrix elements $\langle
n|\rho|n\rangle$ in  a stationary state. Given an initial
condition for the density operator, the evolution of $\rho$,
however, is more useful than the stationary solution. In contrast
with the solution in P-representation[4], in what follows, we
present a solution of the master equation in a number state (Fock
state) basis.

Comparing the above master equation with the general form eq.(2),
we see that $H=(\omega_f-i\kappa/2)a^\dagger a$ and $L=a$. The
auxiliary system is another single-mode field whose annihilation
and creation operators are denoted as  $b$ and $b^{\dagger}$
respectively. By definitions eqs.(7-8) we have
$H_A=(\omega_f+i\kappa/2)b^\dagger b$ and $L_A=b$. Hence the
effective Hamiltonian governing the time evolution of the
bipartite system is
\begin{equation}
\tilde{H}=(\omega_f-\frac{i\kappa}{2})a^{\dagger}a-(\omega_f+\frac{i\kappa}{2})b^{\dagger}b+
i\kappa ab,
\end{equation}
The time evolution operator $U(t)$ corresponding the effective
Hamiltonian (19) reads
\begin{equation}
U(t)=e^{-i(\omega_f-\frac{i\kappa}{2})a^{\dagger}at}
e^{i(\omega_f+\frac{i\kappa}{2})b^{\dagger}b t}e^{g_tab},
\end{equation}
where $g_t=1-e^{-\kappa t}$. We consider a general initial state
$\rho(0)= \sum_{mn}\rho_{mn}|m\rangle\langle n|,$ where
$\rho_{mn}=\langle m|\rho(0)|n\rangle$ is the element of the
density operator at $t=0$.  Under the Fock state basis of the
bipartite system, since $|\psi_{\rho}(0)\rangle=
\sum_{m,n}\rho_{mn}|m\rangle|n\rangle_b,$ with $|n\rangle_b$
satisfying $b^{\dagger}b|n\rangle_b=n|n\rangle_b$ we arrive at
\begin{eqnarray}
\langle p|\rho(t)|q\rangle
&=&e^{-i\omega_f(p-q)t-\kappa(p+q)t}\times\cr
&&\times\sum_{m=0}^\infty
\frac{\sqrt{(p+m)!(q+m)!}}{(m!)^2}{g_t^m}\rho_{p+m,q+m}.
\end{eqnarray}
It is obvious that all elements of the density operator are
damped-oscillation function of time, the decay rate $\kappa(p+q)$
depends on the sum of $p$ and $q$, which are initial
condition-independent. Furthermore, we notice that the diagonal
elements of the density operator, which represent the population
in the corresponding state decay with different damped rate, the
higher the state, the faster the decay.  When the single-mode is
initially at a thermal state $\rho(0)=e^{-\beta a^\dagger a}$ with
the partition function as its normalization, we find that after
time $t$ the system is still at a thermal state
$\rho(t)=e^{-\beta(t) a^\dagger a}$ with
\begin{equation}
\beta(t)=\beta+\kappa t+\ln(1-e^{-\beta}(1-e^{-\kappa t})).
\end{equation}

Based on equation (21), for the bipartite system in the initial
state corresponding to $\rho(0)$ we have
 \begin{eqnarray}
 \rho(t)&=&\sum_{m=0}^\infty A_m(t)\rho(0) A^\dagger_m(t)\cr
 A_m(t)&=&\sqrt{\frac{g_t^{m}}{m!}}e^{-(i\omega_f+\kappa/2)a^\dagger
 at}a^m,
 \end{eqnarray}
which shows explicitly that the evolution is completely positive
and trace preserving. Therefore there exist an environment and an
evolution $U(t)$ of the system and the environment such that
\begin{equation}
\rho(t)=\mbox{Tr}_E(U(t)(\rho(0)\otimes |0\rangle\langle 0|_E)
U^\dagger(t))
\end{equation}
satisfies the master equation. It should be emphasized that the
environment is exact. For the simple case of a single-mode field
in lossy cavity, the environment is another single-mode field and
the evolution is
\begin{equation}
U(t)=e^{-i\omega_ft a^\dagger a}e^{\theta_t(a^\dagger b-b^\dagger
a)}
\end{equation}
with $\cos\theta_t=e^{-\kappa t/2}$. In fact $A_m(t)=\langle
m_b|U(t)|0_b\rangle$. This provides another Hamiltonian approach
to this kind of problem: two-mode field interacting with one
another with a time-dependent interaction.

In summary, we propose a method to approach solving the master
equation exactly. For this end, we first of all introduce an
auxiliary system, which has the same dimension as the system that
we are interested in. The original system and the ancilla interact
on each other, and the Hamiltonian which governs the time
evolution of the composite system are determined by the master
equation. In this sense, the solution of the master equation might
be computed through the Schr\"odinger equation of the composed
system. We would like to note that whether we can obtain the
solution of the master equation in an explicit form or not depends
on the form of the effective Hamiltonian, but the method presented
here provides a new approach to the exact solution of the master
equation.

This work was supported by Chinese
postdoctoral Fund via Institute of Theoretical Physics, Academia
Sinica, and by K. C. Wong Education Foundation, Hong Kong.

\end{document}